\documentstyle[epsfig]{elsart}
\begin{document}
\begin{frontmatter}
\title{ K$^-$ Meson Production in the Proton--Proton Reaction at 3.67~GeV/c}

\author[Tor]{F.~Balestra} 
\author[LNS]{Y.~Bedfer}
\author[LNS,Tor]{R.~Bertini}
\author[IUCF]{L.C.~Bland}
\author[Gie]{A.~Brenschede\thanksref{aa}}  
             \thanks[aa]{current address: Brokat Infosystems AG - Stuttgart}
\author[LNS]{F.~Brochard\thanksref{ab}}
             \thanks[ab]{current address: LPNHE X, Ecole 
					  Polytechnique Palaiseau}
\author[Tor]{M.P.~Bussa}
\author[JINR]{V.~Tchalyshev}
\author[IUCF]{Seonho~Choi\thanksref{bb}}    
              \thanks[bb]{current address: Temple University, Philadelphia} 
\author[Kr1]{M.~Debowski\thanksref{bc}}   
              \thanks[bc]{current address: FZ-Rossendorf} 
\author[IUCF]{M.~Dzemidzic\thanksref{cc}}  
            \thanks[cc]{current address: IU School of Medicine - Indianapolis} 
\author[LNS]{J.-Cl.~Faivre} 
\author[JINR]{I.V.~Falomkin} 
\author[UPO]{L.~Fava}
\author[Tor]{L.~Ferrero} 
\author[Kr2,Kr1]{J.~Foryciarz\thanksref{dd}} 
      \thanks[dd]{current address: Motorola Polska Software Center - Krakow}
\author[Gie]{I.~Fr\"ohlich}
\author[JINR]{V.~Frolov}
\author[Tor]{R.~Garfagnini} 
\author[TRI]{D.~Gill} 
\author[Tor]{A.~Grasso}
\author[Ros]{E.~Grosse}
\author[LNS]{S.~Heinz}
\author[JINR]{V.V.~Ivanov}
\author[IUCF]{W.W.~Jacobs} 
\author[Gie]{W.~K\"uhn}
\author[Tor]{A.~Maggiora} 
\author[Tor]{M.~Maggiora} 
\author[LNS,Tor]{A.~Manara}
\author[UPO]{D.~Panzieri} 
\author[Gie]{H.-W.~Pfaff} 
\author[Tor]{G.~Piragino}
\author[JINR]{G.B.~Pontecorvo} 
\author[JINR]{A.~Popov}
\author[Gie]{J.~Ritman}
\author[Kr1]{P.~Salabura}
\author[Tor]{F.~Tosello} 
\author[IUCF]{S.E.~Vigdor} 
\author[Tor]{ G.~Zosi}

\address[JINR]{JINR, Dubna, Russia }
\address[IUCF]{Indiana University Cyclotron Facility, Bloomington, Indiana, U.S.A.}
\address[LNS]{Laboratoire National Saturne, CEA Saclay, France}     
\address[Tor]{Dipartimento di Fisica ``A. Avogadro" and INFN - Torino, Italy}
\address[UPO]{Universita' del Piemonte Orientale and INFN - Torino, Italy}
\address[Kr1]{M. Smoluchowski Institute of Physics, Jagellonian University, Krak\'ow, Poland}
\address[Kr2]{H.Niewodniczanski Institute of Nuclear Physics, Krak\'ow, Poland}
\address[Gie]{II. Physikalisches Institut,  University of Gie{\ss}en, Germany}
\address[Ros]{Forschungszentrum Rossendorf, Dresden, Germany}
\address[TRI]{TRIUMF - Vancouver, Canada}

\collaboration{DISTO Collaboration}

\begin{abstract}

  The total cross section of the reaction $pp\rightarrow ppK^+K^-$ has
been determined for proton--proton reactions with $p_{beam}=3.67~
GeV/c$.  This represents the first cross section measurement of the
$pp \rightarrow ppK^-K^+$ channel near threshold, and is equivalent to
the inclusive $pp\rightarrow ppK^-X$ cross section at this beam
momentum.  The cross section determined at this beam momentum is about
a factor 20 lower than that for inclusive $pp\rightarrow ppK^+X$ meson
production at the same CM energy above the corresponding threshold.
This large difference in the $K^+$ and $K^-$ meson inclusive
production cross sections in proton-proton reactions is in strong
contrast to cross sections measured in sub-threshold heavy ion
collisions, which are similar in magnitude at the same energy per
nucleon below the respective thresholds.

\end{abstract}


\end{frontmatter}

\section{INTRODUCTION}

Currently there is much interest to determine the total $K^-$ meson
production cross section in nucleon-nucleon reactions near threshold.
This cross section is of particular importance in heavy-ion physics.
In heavy ion collisions it has been observed that the inclusive
production cross sections for $K^-$ and $K^+$ mesons are nearly equal
in sub--threshold reactions when measured at the same energy per
nucleon below the production thresholds for the reactions
$pp\rightarrow K^-X$ and $pp\rightarrow K^+$X,
respectively~\cite{Bar97}.  This result is surprising for several
reasons. First, the $K^-$ production cross section in near--threshold
proton--proton reactions is expected~\cite{Ef94} to be lower by over
an order of magnitude than the $K^+$ cross section at the same
distance from their respective thresholds.  Second, antikaons ($K^-$,
$\overline{K}^0$) should have a higher absorption probability in heavy
ion collisions than kaons ($K^+$, ${K}^0$), primarily as a result of
strangeness exchange reactions (e.g. $K^- N \rightarrow Y \pi$ with $Y
= \Lambda, \Sigma$).  Although charge exchange reactions (e.g. $K^+ n
\rightarrow K^0 p$) may occur, changes to the observed yield of a
particular charge state will be largely compensated by the
corresponding reverse process.
In this context it has been shown in detailed model calculations that
multi-step processes can not explain the enhanced $K^-$ yield in heavy
ion reactions~\cite{Cas97}.  A promising explanation of this
discrepancy comes from various theoretical models~\cite{kpots} which
suggest that, as a result of partial chiral symmetry restoration in a
dense hadronic medium, antikaons are subject to strongly attractive,
and kaons to slightly repulsive, forces.  These effective interactions
would lower the apparent $K^-$ production threshold in a dense
hadronic medium, and thus enhance the $K^-$ yield in heavy ion
collisions.

However, the comparison of heavy ion with nucleon-nucleon collisions
has relied on some assumptions because, in the latter case, the antikaon 
data were either deduced from
$pp \rightarrow ppK_0\overline{K}_0$ results, or
taken as the sum of different exclusive channels with one or two pions
in the final state~\cite{Ef94}.  Furthermore, apart from a preliminary
upper limit~\cite{Mos98}, there is a lack of pp
antikaon production data in the regime of available energy 
most relevant
in near--threshold heavy ion collisions
(i.e. $\sqrt{s}-\sqrt{s_0} < 0.5~GeV$, where $\sqrt{s_0}$ is the
threshold of the particular channel under consideration)~\cite{Li94}.
The result presented in this work represents the first cross section
determination of the exclusive channel $pp \rightarrow ppK^+K^-$, and is
equivalent to the inclusive $K^-$ production cross section since no
other purely hadronic channels including $K^-$ are kinematically allowed 
at the beam momentum of this experiment
(i.e. $\sqrt{s}-\sqrt{s_0} = 111~MeV <M_{\pi^\circ}c^2$).

\section{EXPERIMENTAL PROCEDURE}

\subsection{Apparatus}

 A proton beam from the SATURNE proton synchrotron with momentum
$p_{beam}=3.67~GeV/c$ was directed onto a liquid hydrogen target of
2~cm length.  All events with at least four charged particles in the
final state were measured with the DISTO spectrometer, which is
described in detail elsewhere~\cite{disto}.  Charged particles were
tracked through a magnetic spectrometer which consisted of a dipole
magnet ($\sim 1.0~T\cdot m$), two sets of scintillating fiber
hodoscopes inside the field and 2 sets of multi-wire proportional
chambers (MWPC) outside the field.  Furthermore, two arrays of
scintillator hodoscopes and water \v Cerenkov detectors were located
behind the MWPCs.  Particle identification was performed using the
correlation between the particle momenta and the \v Cerenkov light
output (see also ~\cite{disto}).  The large acceptance of the
spectrometer ($\approx\pm 15^\circ$ vertical, $\approx\pm 48^\circ$
horizontal in the laboratory reference frame) guarantees a sizeable
efficiency for coincident detection of four charged particles.  The
measurement of all particles in the final state allowed 4-momentum
conservation to be used in addition to particle identification for
effective background suppression in order to identify the $ppK^+K^-$
final state.

\subsection{Data Selection}

\begin{figure}[tht]
    \centering
    \mbox{\epsfig{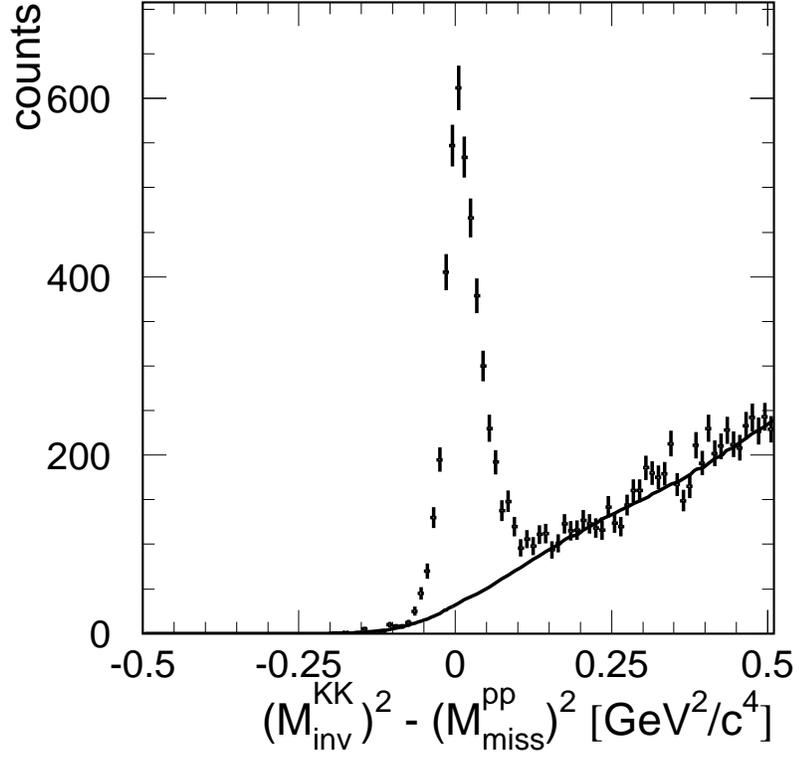}}
    \caption[]{Distribution of $(M_{inv}^{KK})^2 - (M_{miss}^{pp})^2$. The
               data points are for events with kaon identification and
               the solid histogram is the scaled background deduced from
               events without kaon identification.}
    \label{fig:I-M}
\end{figure}

Since the 4-momenta of all particles in the final state were measured,
the events are kinematically over--determined.  Therefore, 4-momentum
conservation can be used for a substantial background suppression by
requiring that the proton-proton missing mass ($M_{miss}^{pp}$) be
equal to the $K^+K^-$ invariant mass ($M_{inv}^{KK}$). The
distribution of $(M_{inv}^{KK})^2-(M_{miss}^{pp})^2$ is plotted in
Figure~\ref{fig:I-M}.  The peak near
$(M_{inv}^{KK})^2-(M_{miss}^{pp})^2 = 0$ results from events where the
$ppK^+K^-$ event hypothesis was correct, and is the basis for the
cross section values quoted below.  This peak is superimposed on a
background resulting from imperfect $\pi-K$ separation in the \v
Cerenkov detectors in a small fraction of events of the type
$pp\rightarrow pK^+\Lambda \rightarrow ppK^+\pi^-$ or $pp \rightarrow
pp\pi^+\pi^-X$.  An estimate of the background underneath the
$ppK^+K^-$ peak is given by the solid curve in
Fig.~\ref{fig:I-M}. This estimate was determined by scaling the
$(M_{inv}^{KK})^2-(M_{miss}^{pp})^2$ distribution, measured for all
events before requiring kaon identification, by a factor 0.002
in order to match the data in Fig.~\ref{fig:I-M}
above $0.15~GeV^2/c^4$.  From this histogram it is determined that the
background accounts for about 13\% of the yield in Fig.~\ref{fig:I-M}
with $|(M_{inv}^{KK})^2-(M_{miss}^{pp})^2| < 0.09~GeV^2/c^4$. This
background has been subtracted in the subsequent analysis as described
below.  To extract a total cross section from the data shown in
Fig.~\ref{fig:I-M}, a correction for the detector acceptance and an
absolute normalization must also be determined.

\subsection{Acceptance Corrections}

  The correction of the measured yields for the detector acceptance
has been evaluated by means of Monte Carlo simulations, which after
digitization of the simulated detector hits, were processed through
the same analysis chain as the measured data.  The detector acceptance
was determined as a multi-dimensional function of the relevant
kinematic degrees of freedom of the particles in the final state.
After accounting for the azimuthal and reflection symmetries, the
detector acceptance was non-zero over the full kinematically allowed
region.  Thus, the acceptance correction is essentially independent of
the actual phase space distribution of the final state, and has been
determined assuming a uniform phase space distribution in the
simulations.  This has been verified by calculating the acceptance
correction matrix using different initial distributions in the
simulations, and observing that the measured yield varied less than
the systematic error associated to the acceptance correction.
%
%
Although eight linearly independent degrees of freedom are in
principle required to fully describe the $pp\rightarrow ppX
\rightarrow ppK^+K^-$ reaction, a five dimensional acceptance
correction matrix is sufficient, in part because the cross section
cannot depend upon the azimuthal orientation of the event.
Furthermore, we have compared the observed angular distribution
of the $X\rightarrow K^+K^-$ decay with that from simulations.  The
comparison of data with simulations has a $\chi ^2/n=1.4$ for a
S-wave distribution and a $\chi ^2/n=17.6$ for a P-wave
distribution.  Therefore, in the simulations uniform distributions
with respect to these three angular variables were integrated over
when determining the acceptance as a function of the remaining five
variables. Finally, the raw data were corrected on an event-by-event
basis, via a weighting factor determined from the simulated acceptance
function for the appropriate kinematic bin.

 Since the acceptance varies as a function of the kinematic
distribution of the final state, one can not simply reduce the
acceptance to a single number without making assumptions about the
phase space distribution of the particles in the final state.
Nevertheless, a reasonable estimate of the average acceptance can be
determined for kaon pairs with an invariant mass equal to the phi
meson mass (where most of the measured kaon pairs are observed) and an
isotropic distribution in the other kinematic variables.  In this case
the product of the geometrical acceptance times the tracking
reconstruction efficiency is 21.4\%. 

\subsection{Background Subtraction}

\begin{figure}[tht]
    \centering 
    \mbox{\epsfig{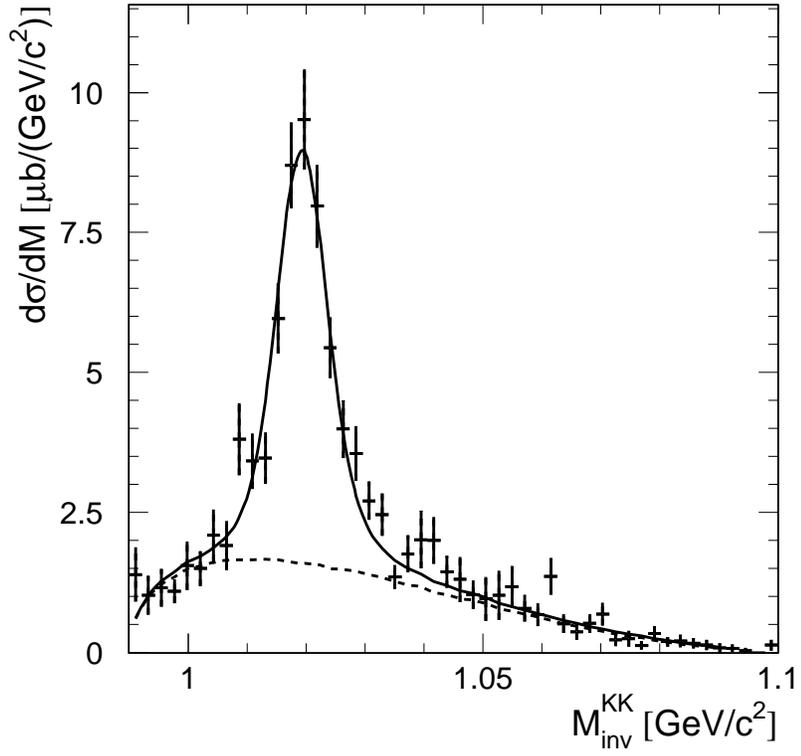}}
    \caption[]{Spectrum of $M_{inv}^{KK}$ after acceptance correction
               and background subtraction.  The solid curve is a fit to
               the data with the sum of a non-resonant $K^+K^-$
               contribution (dashed curve) and the natural line-shape
               of the $\phi$ resonance folded with the detector
               resolution.}
    \label{fig:Minv}
\end{figure}

 The background contribution from non-$ppK^+K^-$ events, shown as the
solid curve in Fig.~\ref{fig:I-M}, must be subtracted from the data in
order to determine the $K^+K^-$ yield.  The subtraction was performed
on the $M_{inv}^{KK}$ distribution.  For this, the $M_{inv}^{KK}$
distribution for the background events was determined by applying the
acceptance correction matrix (determined for simulated $pp \rightarrow
ppK^+K^-$ reactions) to events subjected to the same kinematic
requirement $|(M_{inv}^{KK})^2-(M_{miss}^{pp})^2| < 0.09~GeV^2/c^4$,
but not to the kaon identification conditions with the \v Cerenkov
detectors.  The resulting distribution was scaled by the same
factor (0.002) used by the solid curve in Fig.~\ref{fig:I-M}, and then
subtracted from the acceptance corrected $M_{inv}^{KK}$ distribution
that included both $ppK^+K^-$ and background events.  Finally, the
$M_{inv}^{KK}$ spectrum, after full acceptance corrections and
background subtraction, is shown in Figure~\ref{fig:Minv}.  The curves
are fits to the data as described below.

\subsection{Absolute Normalization}
\label{absnorm}

The absolute normalization of the $ppK^+K^-$ cross section was
determined by measuring the yield relative to that of a {\em
simultaneously} measured channel with known cross section.  For this
work the reference channel was the reaction $pp\rightarrow pp\eta$ for
which a large amount of data exist~\cite{eta}.  This method to
determine the absolute normalization was chosen because it greatly
reduced the large systematic uncertainty associated with the absolute
calibrations of both beam intensity and absolute trigger
efficiency. In order to provide the absolute cross section calibration
the existing published data was first interpolated to get the $\eta$
production cross-section at the beam momentum of the present
measurement, and then the appropriate acceptance corrections were
applied to determine the $K^+K^-/\eta$ total cross section ratio from
our yields.

  The cross section of the reaction $pp\rightarrow pp\eta$ has been
interpolated to our beam momentum with several parameterizations that
vary smoothly with beam momentum.  From these interpolations we
estimate the exclusive $pp \rightarrow pp\eta$ production cross
section to be $135\pm 35\mu b$ at $p_{beam}=3.67~GeV/c$.  The
parameterizations describe the existing $\eta$ production data well
with the exception of the single measurement at $p_{beam}=2.8~GeV/c$
by E.~Pickup et al.~\cite{Pic62} which is significantly
underestimated.  This discrepancy has been neglected since that
measurement is subject to a large systematic error associated with a
quite substantial background subtraction.

  In the present data, the $\eta$ meson has been identified in the
$M_{miss}^{pp}$ distribution for $pp\pi^+\pi^-X$ events, after
requiring that the four particle missing mass be consistent with
$M_{\pi^\circ}$ as shown in Ref.~\cite{prl}.  The acceptance
correction of the $pp\rightarrow pp\eta \rightarrow pp\pi^+\pi^-\pi^0$
channel was performed similarly to the $pp\rightarrow ppK^+K^-$
discussed above.  The acceptance correction matrix for the $pp\eta$
channel has four dimensions.  These are sufficient to completely
describe this 5 body final state because the full set of 15 kinematic
degrees of freedom (dof) is reduced by four-momentum conservation (-4
dof), the azimuthal symmetry of the event (-1 dof), the requirement
that $M_{miss}^{pp} = M_\eta$ (-1 dof), the isotropic orientation of
the $\eta$ meson decay plane (-3 dof), and the known matrix element
(-2 dof) for the $\eta \rightarrow \pi^+\pi^-\pi^0$
decay~\cite{Ams95}.

\section{Results}

\subsection{Total $K^+K^-$ Cross Section}

 After applying the full acceptance corrections to the $pp\rightarrow
ppK^+K^-$ and $pp\rightarrow pp\eta \rightarrow pp\pi^+\pi^-\pi^0$
channels, the $K^+K^-/\eta$ total cross section ratio is determined to
be $(1.5\pm0.1(stat.)\pm0.4(sys.))\times 10^{-3}$.  The systematic
error quoted here arises from the quadratic sum of the uncertainties
on the $\eta$ and $K^+K^-$ background subtractions (15\% and 5\%
respectively), relative acceptance correction (11\%), trigger bias
(10\%), tracking efficiency (10\%), and the \v Cerenkov particle
identification efficiency (14\%).  This value has also been corrected
for systematic bias effects due to the different scintillating
fiber efficiencies for pions and kaons (-7.5\%) and from the
subtraction of events arising from the target envelope (+5\%).  Based
on the $\eta$ cross section estimated above in Sec.~\ref{absnorm}, a
total cross section for the reaction $pp\rightarrow ppK^+K^-$ of
$(0.20\pm0.011\pm0.08) \mu b$ has been determined, where the second
error quoted here is the quadratic sum of the systematic uncertainty
in the measured yield ratio and the absolute normalization uncertainty
for the $\eta$ production cross section.

\begin{figure}[tht]
    \centering 
    \mbox{\epsfig{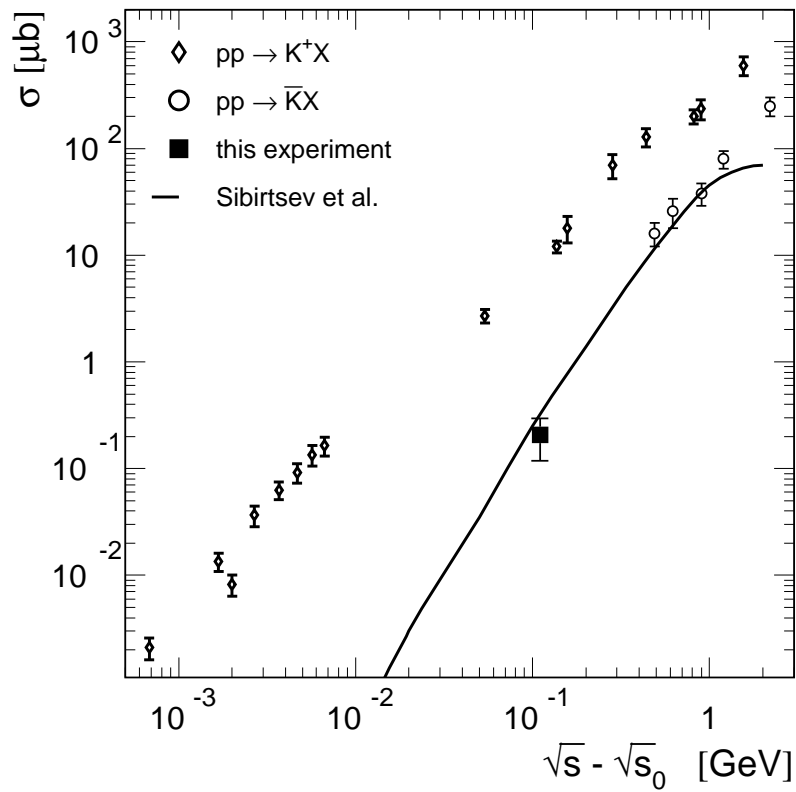}}
    \caption[]{Total kaon and antikaon production cross sections as a
               function of the available energy above the appropriate
               threshold.  The present measurement is the solid data
               point and the open circles are antikaon points deduced
               from the literature. The curve is a model prediction
               described in the text and the open diamond points are
               cross section values for positive kaon production.}
    \label{fig:xc}
\end{figure}

  The present total $pp\rightarrow ppK^+K^-$ cross section value is
plotted as the solid data point in Figure~\ref{fig:xc}, where it is
compared with estimates of $K^-$ inclusive production cross sections
at higher energy (open circles) taken from the
literature~\cite{Ef94}. These additional data points have been deduced
from other reactions either by assuming $\sigma_{pp\rightarrow K^-X} =
\sigma_{pp\rightarrow \overline{K}^0X}$, or by taking the sum of
exclusive channels with one or two pions in the final state.  The
solid curve, corresponding to a prediction from Sibirtsev et
al.~\cite{Sib97} using a one meson exchange model including $\pi, K,$
and $K^*$ mesons, accounts well for the $K^-$ cross section measured
in this experiment near the production threshold.  In comparison, the
total cross sections for inclusive $K^+$ production shown as the open
diamonds~\cite{k+} in Fig.~\ref{fig:xc}, are more than an order of
magnitude larger at comparable distances above the $K^+$ threshold.

\subsection{$\phi$ Meson}

 Near threshold the $K^-$ meson is produced, to a large degree,
by the decay of the
$\phi$ meson as an intermediate state~\cite{prl}. The fraction of the cross
section from the resonant production can be determined from the
$M_{inv}^{KK}$ distribution.   The $M_{inv}^{KK}$ spectrum in 
Fig.~\ref{fig:Minv} has been fit
with the sum of a non-resonant contribution and a peak from the $\phi$
resonance.  The shape of the non-resonant contribution was assumed to
be given by the $M_{inv}^{KK}$ distribution for an ensemble of
events that are uniformly distributed according to four body
($ppK^+K^-$) phase space.  The shape of the $\phi$ resonance was given
by the natural line-shape folded with a Gauss function to account for
the detector resolution.  When treating the width ($\sigma$) of the 
Gaussian as a free parameter, we find an optimal fit with
$\sigma=3.5\pm0.5~MeV/c^2$, in good agreement with simulations of 
the detector performance.

  The total $K^+K^-$ cross section value, as well as the resonant and
non-resonant components, are summarized in Table~\ref{tab:crosssec}.
 After correction for the corresponding partial width~\cite{PDG},
the total $\phi$ meson production cross section is $0.19\pm 0.014 \pm
0.08$ $\mu b$.  The systematic errors quoted have the same meaning as
explained above.

\begin{table}[hhh] 
 \caption[]{Total exclusive production cross section for the reaction
   $pp\rightarrow ppK^+K^-$ at 3.67 GeV/c and for the resonant ($\phi$
   meson) and non-resonant contributions.  }

 \begin{center}
  \begin{tabular}{cl}
   \hline \hline 
   Meson Species           &  Cross Section [$\mu b$]  \\ 

   total $K^+K^-$          &  $0.20 \pm 0.011  \pm 0.08 $   \\


   $\phi \rightarrow K^+K^- $ 
                           &  $0.09 \pm 0.007 \pm 0.04$   \\ 

   non-resonant  $K^+K^-$  &  $0.11 \pm 0.009  \pm 0.046$   \\
   \hline \hline
  \end{tabular}
 \end{center}
 \label{tab:crosssec}
\end{table}

\section{SUMMARY}

  In conclusion, the total cross section of the reaction
$pp\rightarrow ppK^+K^-$ has been determined for
$p_{beam}=3.67~GeV/c$.  This is the first cross section measurement of
the $pp\rightarrow ppK^+K^-$ channel near threshold, and is equivalent
to the inclusive cross section at this beam momentum.  The cross
section determined here is more than a factor 20 lower than the
measured~\cite{k+} and calculated~\cite{Sib95} $pp\rightarrow K^+ + X$
cross section at the same CM energy above threshold.  This large
difference in the $K^+$ and $K^-$ meson production cross sections in
proton-proton collisions is in strong contrast to the nearly equal
cross sections measured in sub-threshold heavy ion collisions at the
same distance from the respective thresholds.  Since this discrepancy
between heavy ion collisions and proton-proton reactions has been
interpreted as possible evidence for in--medium modifications of the
$K^-$--nucleon interaction, it would be very useful to study the
evolution of the relative $K^+$ and $K^-$ meson yields at equal
energies from the threshold versus increasing mass number of the
colliding nuclei.

 {\it {ACKNOWLEDGMENTS}}

This work has been supported in part by the following agencies:
CNRS-IN2P3, CEA-DSM, NSF, INFN, KBN (2 P03B 117 10 and 2 P03B 115 15)
and GSI.

\newpage

\end{document}